\documentclass[journal,onecolumn,draftcls,12pt]{IEEEtran}
\IEEEoverridecommandlockouts
\usepackage[pdftex]{graphicx}
 \usepackage{lipsum,graphicx}
\usepackage{float}
\usepackage{amsmath,amssymb,amsfonts}
\usepackage{textcomp}
\usepackage{url}
\usepackage{algorithm}
\usepackage{algpseudocode}
\usepackage[T1]{fontenc}% optional T1 font encoding
\usepackage{mathtools, cuted}
\usepackage{lipsum, color}
\usepackage{amsthm}

\usepackage{mathtools}
\usepackage{graphics}

\usepackage{algpseudocode}
\usepackage{float}
\usepackage{verbatim}
\usepackage{array}
\usepackage{cite}
\usepackage{lipsum}
\usepackage{mathtools}
\usepackage{icomma}
\usepackage{optidef}
\usepackage[font=footnotesize]{subcaption}
\usepackage[font=footnotesize]{caption}
\usepackage{mwe}
\usepackage{textgreek}
\usepackage{tabularx}
\usepackage{multirow}
\usepackage{stackengine}

\begin{document}
\title{Reconfigurable Intelligent Surface-Assisted mmWave multi-UAV Wireless Cellular Networks}
\author{
    \IEEEauthorblockN{Lisi Jiang and Hamid Jafarkhani}\\

  %  \IEEEauthorblockA{\IEEEauthorrefmark{7}{\small{Department of Electrical Engineering,   University of South Carolina, 
   % \{matolak\}@cec.sc.edu}}}\\

 %\IEEEauthorblockA{\IEEEauthorrefmark{2}{\small{School of Electronic Engineering and Computer Science,   Queen Mary University of London,   \{c.pan,  maged.elkashlan\}@qmul.ac.uk}}}
\thanks{This work was presented in part at the 2021 IEEE International Conference on Communications \cite{ICC2021}. This work was supported in part by NSF Awards CCF-2008786 and ECCS-1642536. L. Jiang and H. Jafarkhani are with Center for Pervasive Communications and Computing, University of California, Irvine, Irvine, CA 92697, USA (e-mail: $\{$lisi.jiang, hamidj$\}$@uci.edu). 
}
}
\maketitle
\begin{abstract}
Unmanned aerial vehicles (UAVs) have brought a lot of flexibility in the network deployment. However, UAVs suffer from the high mobility and instability. To improve the capacity and reliability of the UAV networks, millimeter-wave (mmWave) and reconfigurable intelligent surfaces (RISs) can be used in the system. In this paper, we consider  an RIS-assisted mmWave UAV  wireless cellular network, where UAVs serve several users with the help of multiple RISs. We jointly optimize the deployment, user scheduling, beamforming vector, and RIS phases to maximize the sum-rate, with the constraints of the minimum rate, the UAV movement, the analog beamforming, and the RIS phases.  To solve this complex problem, we use an iterative method, in which when we optimize one variable, we fix the other three variables. When optimizing the deployment, we find the optimal position for the UAV by a sphere search. Then, we formulate a mixed-integer non-linear problem (MINLP) to find the best scheduling. A spatial branch-and-bound (sBnB) method is used to solve the MINLP. When Optimizing the beamforming vector and the RIS phases, we propose an iterative algorithm that relies on the equivalence between the  maximization of the sum-rate and the minimization of the summation of weighted mean-square errors (sum-WMMSE). The majority-minimization method is used to deal with the constant-modulus constraints for the analog beamforming and RIS phases. The proposed joint optimization offers significant advantages over the system without beamforming and RIS phase optimization and the system without deployment optimization. In addition, the RIS can compensate for the loss of throughput due to the blockage, especially in low flight altitudes.
\end{abstract}
\newpage
\section{Introduction}
 High capacity, low latency, and ultra reliability are required for the future wireless systems. To satisfy these requirements, the wireless networks will evolve into ultra dense distributed cooperating \cite{Koyuncu2012} and self-organized networks \cite{wang2014challenges} that can handle interference using multiuser decoding capabilities \cite{Kazemitabar2009MultiuserAntennas}. To realize the evolution of the wireless networks, unmanned aerial vehicles (UAVs) are introduced into the future wireless networks because of their flexibility, mobility, and fast deployment \cite{xiao2016enabling,zhang2019survey}. Utilizing UAVs in scenarios such as wireless sensor networks (WSNs) \cite{guo2020optimal,koyuncu2018deployment,guo2016sensor}, caching aided wireless networks \cite{chen2017caching}, cloud radio access networks (CRANs) \cite{ma2021concise}, and cellular networks will bring great improvement in the system coverage and flexibility.  Among these scenarios, the UAV-assisted wireless cellular network is a promising technology to enable significantly enhanced UAV-ground communications \cite{zeng2018cellular}. In UAV-assisted wireless cellular networks, a UAV can serve as a flying base-station (BS), an aerial radio access point, and an aerial relay to expand wireless coverage and provide data transmission towards physical objects. 

Although UAVs bring a lot of flexibility in deploying the network, their high mobility and instability severely impair the quality of communication. To improve the reliability and capacity, technologies such as millimeter wave (mmWave),  reconfigurable antennas \cite{cetiner2004multifunctional,cetiner2008patent,grau2008reconfigurable,fazel2009space}, or reconfigurable intelligent surface (RIS) \cite{wu2018intelligent,huang2019reconfigurable} can be used in UAV networks. MmWave communications can provide large bandwidth to support high-rate communication in UAV networks\cite{xiao2016enabling,xiao2019unmanned}. Reconfigurable antennas and RISs can improve the reliability of the system by changing the wireless scattering environment. Both reconfigurable antennas \cite{almasi2018new,he2018low} and RISs \cite{wang2020intelligent,yang2020mimo,wang2020joint} can intelligently configure the mmWave wireless environment to improve the communication quality between the transmitter and receiver. However, for UAV networks, RISs are preferred since they use passive units rather than active units, which only result in a signal phase shift without power consumption. Also, RIS can help improve the channel quality when the line-of-sight (LoS) path is affected by physical obstacles or under harsh weather environments, e.g., rain. 

There have been several papers in the literature on the mmWave UAV network or RIS-assisted UAV networks. For example, in \cite{zhong2017research}, a spatial interference channel model is established for UAV groups and the expression of signal to interference plus noise ratio (SINR) is obtained. In \cite{xiao2019unmanned}, a joint optimization of the UAV-BS deployment and beamforming to maximize the achievable sum-rate in a multi-user mmWave-UAV system is proposed. In \cite{hua2020uav}, the UAV-BS link is assisted and optimized by the RIS. In \cite{li2020reconfigurable}, trajectory and beamforming are jointly designed for a scenario in which one UAV serves one user. In the aforementioned papers, either the UAVs are not considered as flying BSs to serve multiple users or the RISs are not implemented in the UAV networks. Furthermore, the mmWave technology has not been implemented in the RIS-assisted UAV networks. To fully exploit the capacity and coverage of the UAV networks, the UAVs should be able to serve multiple users using mmWave beams, and benefit from RISs to improve the reliability.

In this paper, we consider a scenario that includes UAVs as flying BSs to serve users using mmWave beams with the help of multiple RISs. To improve the throughput of the system, we propose an optimization problem, which jointly considers the deployment, user scheduling, beamforming vector, and RIS phases to maximize the sum-rate. We also include constraints on the minimum rate, the movement of the UAV, the analog beamforming, and the RIS phases. To solve this complex problem, we use an iterative method. In our method, we optimize one variable while fixing the other three variables. When optimizing the deployment, we find the optimal position for the UAV by a sphere search. Then, we formulate a mixed-integer non-linear problem (MINLP) to find the best scheduling. A spatial branch-and-bound (sBnB) method is used to solve the MINLP. When optimizing the beamforming vector and the RIS phases, we propose an iterative algorithm by making use of the equivalence between the sum-rate maximization and the minimization of the summation of weighted mean-square errors (sum-WMMSE). The majority-minimization method is used to deal with the constant-modulus constraints for the analog beamforming and RIS phases. The proposed joint optimization outperforms the system without RIS assistance and the system without deployment optimization.

Our contributions are summarized as follows:
\begin{itemize}
\item We propose using RISs and mmWave beams in a network with flying BSs, i.e., UAVs, to improve the capacity, coverage, and reliability of the UAV networks. The results show that our integrated system can provide great gains in terms of sum-rate and minimum rate compared with the existing systems in the literature. Also, the RIS can compensate for the loss of throughput due to the blockage, especially at low flight altitudes.
%\item We  propose  a  new  scenario where  UAVs serve users with the help of multiple RIS.
\item We propose and solve a joint optimization of the UAV deployment,  scheduling, RIS phases, and beamforming to maximize the system throughput. A MINLP is formulated for the scheduling optimization and a sBnB method is proposed to solve it. The solution also uses the equivalence between the sum-rate maximization and the sum-WMMSE to optimize the beamforming vector and RIS phases.
%\item A MINLP is formulated for the scheduling optimization and a sBnB method is proposed to solve it.
%\item We propose a joint beamforming vector and RIS-phase optimization, where we make use of the equivalence between the sum-rate maximization and the sum-WMMSE. The majority-minimization method is used to deal with the constant-modulus constraints for the analog beamforming and RIS phases.
\end{itemize}

The organization of the paper is as follows: Section~\ref{sec:system_model} presents the system model. In Section~\ref{sec:joint_problem}, the joint optimization problem is formulated. Section~\ref{sec:optimization} provides the solution to the joint optimization problem. Numerical results are presented in Section~\ref{sec:simulation}. Section~\ref{sec:conclusion} concludes the paper.    

\emph{Notation:} Small letters, bold letters, and bold capital letters designate scalars, vectors, and matrices, respectively. Matrices $\mathbf{A}^T$ and $\mathbf{A}^{H}$ are the transpose and the Hermite transpose of matrix $\mathbf{A}$, respectively. Operator $\lceil \cdot \rceil$ denotes the ceiling function. 
 \section{System and channel model}
 \label{sec:system_model}
 \subsection{System model}
We consider a multiple rotary-wing UAV, multi-RIS, and multi-user scenario in which $N$ UAVs function as flying BSs to serve $K$ ground users with the help of $R$ deployed RISs on the ground ($R \leq N  \leq K$).  UAVs include $N_t$ antennas, ground users contain a single antenna, and each RIS is equipped with $N_\text{RIS}$ reflecting elements.

The rotary wing UAVs are relatively static over a given geographic area. We assume a quasi-static mobility model. That is within a timeblock, the UAVs are assumed static and between timeblocks, the UAVs can move one small step. Each timeblock includes $M$ timeslots. The users are assumed to move continuously and their locations are collected every $M$ timeslots. We divide the users into $N$ groups for the $N$ UAVs to serve. To guarantee that all users can be served, we assume  $\lceil\frac{K}{N}\rceil \leq M$. 

The goal of our system is to maximize the system throughput in each timeblock. To do this, we need to jointly group the users, design the scheduling, and optimize the UAV beams and RIS phases. In Fig. \ref{Fig_UAV}, we illustrate an example of a 2-UAV, 2-RIS, and 10-user system of different timeblocks. At different timeblocks, the UAVs will serve different users according to the scheduling order. In the example, at Timeblock 1, UAVs 1 and 2 serve 5 users each. At Timeblock 2, UAV 1 serves 4 users and UAV 2 serves 6 users. In this example, at Timeslot 7 of Timeblock 1, UAV 1 serves User 5 with the help of RIS 1 because of the obstacles while UAV 2 serves User 7 with the help of RIS 2 to avoid blockage. Similarly, at Timeslot 2 of Timeblock 2, UAVs 1 and 2, with the help of RISs 1 and 2, respectively, serve Users 4 and 5.
 \begin{figure}[!htb]
     \centering
     \includegraphics[width = 14 cm]{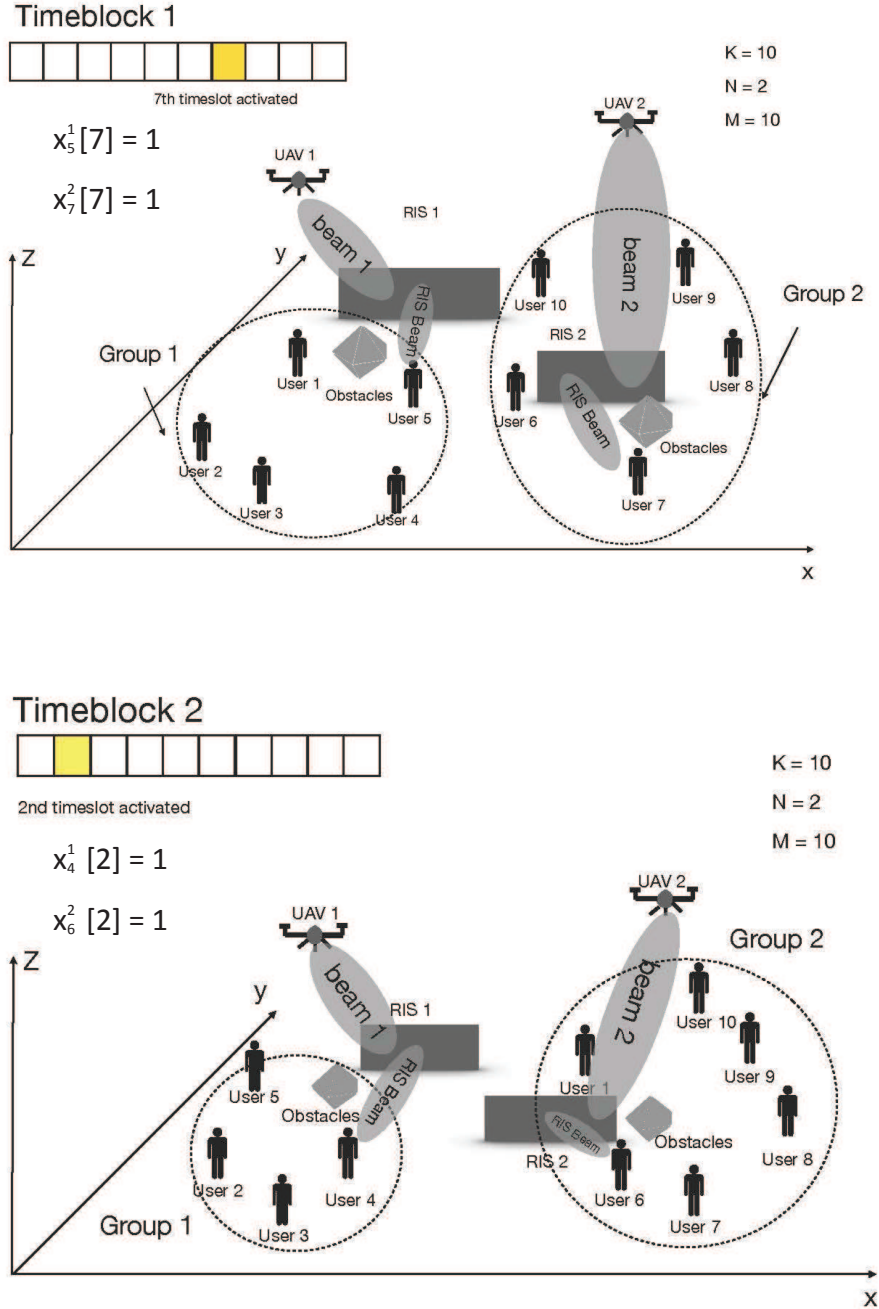}
     \caption{An example of our system model with 2 UAVs, 10 users and 10 timeslots}
     \label{Fig_UAV}
 \end{figure}
 
 \subsection{Channel model}\label{sec:channel}
\subsubsection{UAV-user channel}
\begin{figure}[!htb]
    \centering
    \includegraphics[width = 8 cm]{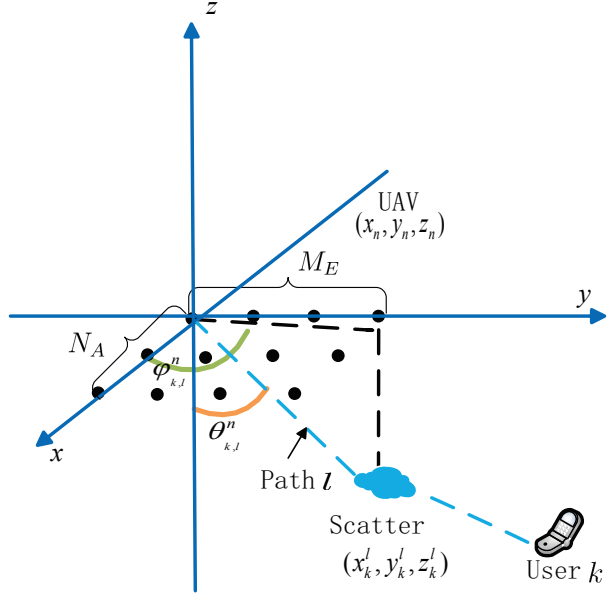}
    \caption{UAV-user channel model illustration with $N_A\times M_E$ antennas at UAV $n$}
    \label{Fig_UAV}
\end{figure}
In our scenario, we assume each UAV is carrying a uniform planar array (UPA) with one RF chain which operates at mmWave band. The widely used extended Saleh-Valenzuela model \cite{saleh1987statistical} is adopted for the channel model. Affected by wireless scattering environment and UAV altitude, a blockage may appear. Under the scenario of  blockage, the channel vector $\mathbf{h}_k^n$ is expressed as
\begin{equation}
\label{Non_LOS}
    \mathbf{h}_k^n =  \sqrt{\frac{N_t}{L_k^n}}\sum_{l=1}^{L_k^n} a_{k,l}^n\mathbf{\alpha}(\theta_{k,l}^n, \varphi_{k,l}^n),
\end{equation}
where $a_{k,l}^n$ is the channel gain coefficient of the $l^{\text{th}}$ cluster from User $k$ to UAV $n$, $\theta_{k,l}^n$ and $\varphi_{k,l}^n$ are the elevation steering angle and azimuth angle of the $l^{\text{th}}$ cluster from User $k$ to UAV $n$, and $L_k^n$ is the total number of clusters for User $k$ to UAV $n$, respectively. $\mathbf{\alpha}(\theta_{k,l}^n)$ is the steering vector function for the UPA. For a $M_E\times N_A$ ($N_t = M_EN_A$) UPA, the steering vector is defined as
\begin{equation}
\label{steering_vec}
\begin{array}{cc}
    \mathbf{\alpha}(\theta_{k,l}^n, \varphi_{k,l}^n)  & = \frac{1}{\sqrt{N_t}}[1,...,e^{j\pi(c-1)\mu_{k,l}^n + (p-1)\nu_{k,l}^n}
     \\
     & ,..., e^{j\pi(N_A-1)\mu_{k,l}^n + (M_E-1)\nu_{k,l}^n}]^T,
\end{array}
\end{equation}
where $\mu_{k,l}^n \triangleq \sin(\theta_{k,l}^n)\cos(\varphi_{k,l}^n)$ and $\nu_{k,l}^n \triangleq \sin(\theta_{k,l}^n)\sin(\varphi_{k,l}^n)$. The steering angles $\theta_{k,l}^n$ and $\varphi_{k,l}^n$ depend on the location of UAV $n$ and User $k$. We denote $(x_n, y_n, z_n)$ as the location of UAV $n$ and $(x_k^l, y_k^l)$ as the location of the $l^{\text{th}}$ scatter for User k. Then, $\theta_{k,l}^n$ and $\varphi_{k,l}^n$ can be calculated as
\begin{equation}
\left\{
\begin{array}{cc}
     &\theta_{k,l}^n = \arctan(\frac{\sqrt{(x_k^l - x_n)^2 + (y_k^l - y_n)^2}}{z_n}),  \\
     &\varphi_{k,l}^n = \arctan(\frac{y_k^l - y_n}{x_k^l - x_n}) - \pi\min(\text{sign}(x_k^l - x_n),0).
\end{array}
\right.
\end{equation}

 Under the scenario of no-blockage, there is an LOS path. Then, we ignore non-LOS paths and the channel becomes
\begin{equation}
\label{LOS}
    \mathbf{h}_k^n =  \sqrt{N_t}a_{k}^n\mathbf{\alpha}(\theta_{k}^n, \varphi_{k}^n),
\end{equation}
where $a_{k}^n$ is the channel gain coefficient of the LOS path from User k to UAV $n$, $\theta_{k}^n$ and $\varphi_{k}^n$ are the elevation steering angle and azimuth angle of the LOS path from User k to UAV $n$, respectively.

The no-blockage probability can be described as a function of the elevation angle $\xi_k^n = \arctan(z_n/D_k)$, where $D_k$ is the horizontal distance from UAV $n$ to User $k$ and $z_n$ is the UAV's altitude \cite{mozaffari2017mobile}. The no-blockage probability is expressed as:
\begin{equation}
\label{plos}
    P_\text{No-Block}(\xi_k^n) = \frac{1}{1 + a\exp(-b(\xi_k^n-a))},
\end{equation}
where $a$ and $b$ are the positive modeling parameters depending on the propagation environment, e.g., rural, urban, or dense urban. The blockage probability can be accordingly calculated by
\begin{equation}
    P_\text{Block} = 1-P_\text{Non-Block}.
\end{equation}
The no-blockage probability increases as the elevation angle increases and it approaches 1 when $z_n$ is large enough. As such, it is more important to have RIS when the altitude is low. This observation is confirmed by simulation results in Section V.

\subsubsection{UAV-RIS-user channel}
We denote the channel between RIS $r$ and UAV $n$ as $\mathbf{G}_r^n$. The channel between RIS $r$ and User $k$ is denoted by $\mathbf{h}_k^r$. We use the same Saleh-Valenzuela model as Eq. (\ref{Non_LOS}) to model  $\mathbf{G}_r^n$ and $\mathbf{h}_k^r$.
For $\mathbf{h}_k^r$, it can be expressed as
\begin{equation}
  \mathbf{h}_k^r =  \sqrt{\frac{N_\text{RIS}}{L_k^r}}\sum_{l=1}^{L_k^r} a_{k,l}^r\mathbf{\alpha}(\theta_{k,l}^r,\varphi_{k,l}^r), 
\end{equation}
where $a_{k,l}^r$ is the channel gain coefficient of the $l^\text{th}$ cluster from RIS $r$ to User $k$ and $\mathbf{\alpha}(\theta_{k,r}^n)$ is the steering vector using the same model as (\ref{steering_vec}). 

For $\mathbf{G}_r^n$, it can be expressed as
\begin{equation}
  \mathbf{G}_r^n =  \sqrt{\frac{N_tN_\text{RIS}}{L_r^n}}\sum_{l=1}^{L_r^n} a_{r,l}^n\mathbf{\mathbf{\alpha}^r(\theta_{r,l}^{n,r},\varphi_{r,l}^{n,r})\alpha}^t(\theta_{r,l}^{n,t},\varphi_{r,l}^{n,t})^H, 
\end{equation}
where $a_{r,l}^n$ is the channel gain coefficient of the $l^\text{th}$ cluster from UAV $n$ to RIS $r$ and $\mathbf{\alpha}^t(\theta_{r,l}^{n,t},\varphi_{r,l}^{n,t})$ and $\mathbf{\alpha}^r(\theta_{r,l}^{n,r},\varphi_{r,l}^{n,r})$ are the transmitting steering vector and receiving steering vector, respectively.

The overall channel between User $k$ and UAV $n$ via RIS $r$ can be expressed as
\begin{equation}
    \mathbf{h}_{k,r}^n = \mathbf{h}_k^r\Theta_r\mathbf{G}_r^{n},
\end{equation}
where $\Theta_r = \text{diag}(e^{j\theta_{r,1}},...,e^{j\theta_{r,N_\text{RIS}}}) $ is the phase-shift matrix of the $r^\text{th}$ RIS and $\theta_{r,m}\in [0,2\pi]$ denotes the phase shift associated with the $m^\text{th}$ passive element of the $r^{th}$ RIS.

\section{Problem formulation}
\label{sec:joint_problem}

\subsection{Scheduling}
The main idea behind scheduling is to cluster users into different groups, one group for each UAV, such that within each group only one user is served by one UAV. Therefore, in each timeblock, we have 3 rules for scheduling: (i) in one timeslot, each UAV can only serve at most one user; (ii) in one timeslot, each user can only be served by at most one UAV; and (iii) across all timeslots, all K users should be scheduled at least once.

To describe the procedure of scheduling, we denote the binary variable $x_k^n[m] \in \{0,1\}$ to indicate whether User $k$ is scheduled to be served by UAV $n$ at Timeslot $m$, i.e.,
\begin{equation}
 x_k^n[m] = \left\{
             \begin{array}{lr}
             1, \text{if User $k$ is scheduled to be served by UAV $n$ at Timeslot $m$}&  \\
             0, \text{otherwise}&  
             \end{array}\right.   
\end{equation}

For Rule (i), we have
\begin{equation}
\label{sche_cons_1}
0\leq\sum_{k=1}^K x_k^n[m] \leq 1.    
\end{equation}

For Rule (ii), we have
\begin{equation}
\label{sche_cons_2}
0\leq\sum_{n=1}^N x_k^n[m] \leq 1.    
\end{equation}

% For Rule iii, we have
% \begin{equation}
% \label{sche_cons_3}
% 0\leq\sum_{n=1}^N\sum_{k=1}^K x_k^n[m] \leq N    
% \end{equation}

For Rule (iii), we have
\begin{equation}
\label{sche_cons_4}
\sum_{m=1}^{M}\sum_{n=1}^N x_k^n[m] \geq 1.   
\end{equation}

\subsection{UAV Beamforming and RIS Reflecting}
 We denote $\mathbf{w}^n_k[m]$ as the beamforming vector from UAV $n$ to User $k$ at Timeslot $m$ with the constant-modulus constraint $|[\mathbf{w}^n_k[m]]_t| = \frac{1}{\sqrt{N_t}}, t = 1,...,N_t$.

Then, at Timeslot m, the received signal from UAV $n$ to User $k$ is
\begin{equation}
\begin{aligned}
    y_k^n[m] &= x_k^n[m]\sqrt{P}(\mathbf{h}_k^{n^H} + \sum_{r=1}^R\mathbf{h}_k^r\Theta_r[m]\mathbf{G}_r^n)\mathbf{w}^n_k[m]\mathbf{s}_k + \\
    & \sum_{n'\neq n, n'= 1}^N\sqrt{P}(\mathbf{h}_k^{n^{'H}}+\sum_{r=1}^R\mathbf{h}_k^r\Theta_r[m]\mathbf{G}_r^{n'}) \mathbf{w}^{n'}_{k'}[m](\sum_{k'\neq k, k'= 1}^{K}x_{k'}^{n'}[m]\mathbf{s}_{k'}) + n_k.
\end{aligned}
\end{equation}

Then, the achievable data rate from User $k$ to UAV $n$ at Timeslot $m$ can be expressed as

\begin{equation}
    R_k^n[m] = \log_2\left(1+\frac{P|(\mathbf{h}_k^{n^H} + \sum_{r=1}^R\mathbf{h}_k^r\Theta_r[m]\mathbf{G}_r^n) \mathbf{w}^n_k[m]|^2x_k^n[m]}{\sum_{n'\neq n, n'= 1}^N(P|(\mathbf{h}_k^{n^{'H}}+\sum_{r=1}^R\mathbf{h}_k^r\Theta_r[m]\mathbf{G}_r^{n'}) \mathbf{w}^{n'}_{k'}[m]|^2\sum_{k'\neq k, k'=1}^{K}x_{k'}^{n'}[m]) + \sigma^2}\right),
\end{equation}
where $\sigma^2$ is the power of Gaussian white noise at User k and P is the total transmission power at UAVs.
Note that RIS phases $\Theta_r$ may change at different timeslots and we use $\Theta_r[m]$ to denote the RIS phase at Timeslot $m$.

\subsection{Joint optimization}
Our goal is to maximize the throughput of the system in each timeblock. Since UAVs cannot rapidly move from one position to another position far away within a short time, we assume that UAVs can only move one small step in each timeblock or stay static. This means $|\mathbf{p}_n - \mathbf{p}^\text{pre}_n| = d\;\text{or}\; 0$, where $\mathbf{p}_n\triangleq (x_n,y_n,z_n)$ is the position of UAV $n$ and $\mathbf{p}^\text{pre}_n$ is the position of UAV $n$ in the previous timeblock. The parameter $d$ is decided by the UAV's energy constraint. To maximize the throughput of the system, considering all required constraints, the following optimization problem is defined:
\begin{maxi!}
{\{(x_n,y_n,z_n)\},\{x_k^n[m]\},\{\mathbf{w}^n_k[m]\},\{\Theta_r[m]\}}{{\qquad \sum_{m=1}^{M}\sum_{k=1}^{K}\sum_{n=1}^N R_k^n[m]} \label{eq:objectiveopt}}{\label{eq:opt}}{}
\addConstraint{(\ref{sche_cons_1})-}{(\ref{sche_cons_4})} {\label{a}}
\addConstraint{|\mathbf{p}_n - \mathbf{p}^\text{pre}_n|}{= d\;\text{or}\; 0}{\label{b}}
\addConstraint{\sum_{m=1}^{M}\sum_{n=1}^N R_k^n[m]}{\geq \gamma_k}{\label{e}}
\addConstraint{|[\mathbf{w}_k^n[m]]_t|}{ = \frac{1}{\sqrt{N_t}}, t = 1,...,N_t}{\label{f}}
\addConstraint{\Theta_r[m]}{ = \text{diag}(e^{j\theta_{r,1,m}},...,e^{j\theta_{r,N_\text{RIS},m}}).}{\label{g}}
\end{maxi!}
Constraint (\ref{e}) guarantees the minimum data rate for each user. Constraint (\ref{f}) is because of analog beamforming and Constraint (\ref{g}) assures that RIS can only adjust a single phase.

\section{Optimal deployment, scheduling, beamforming, and RIS phases}
\label{sec:optimization}
It is intractable to jointly optimize deployment, scheduling, beamforming, and RIS phases at the same time. We propose an algorithm which iterates among deployment, scheduling, beamforming, and RIS phases to improve the throughput of the system at each iteration.

\subsection{Scheduling}
When optimizing the scheduling, we fix the deployment, the beamforming, and the RIS phases. The data rate is expressed as
\begin{equation}
    \tilde{R}_{k}^n[m] = \log_2\left(1+\frac{Pg_{k}^n[m]x_k^n[m]}{\sum_{n'\neq n, n'= 1}^N\sum_{k'\neq k, k'=1}^KPc_{k',k}^{n'}[m]x_{k'}^{n'}[m] + \sigma^2}\right),
\end{equation}
where we define 
\[g_k^n[m]\triangleq |(\mathbf{h}_k^{n^H} + \sum_{r=1}^R\mathbf{h}_k^r\Theta_r[m]\mathbf{G}_r^n) \mathbf{w}^n_k[m]|^2,\] 
\[c_{k',k}^{n'}[m]\triangleq |(\mathbf{h}_k^{n^{'H}}+\sum_{r=1}^R\mathbf{h}_k^r\Theta_r[m]\mathbf{G}_r^{n'}) \mathbf{w}^{n'}_{k'}[m]|^2.\]

Then, the throughput maximization is formulated as
\begin{maxi!}
{\mathbf{x}}{{\qquad \sum_{m=1}^{M}\sum_{k=1}^{K}\sum_{n=1}^{N} \tilde{R}_k^n[m]} \label{eq:objectiveopt_2}}{\label{eq:opt_2}}{}
\addConstraint{(\ref{sche_cons_1})-}{(\ref{sche_cons_4})}{\label{s_2}}
\addConstraint{\sum_{n=1}^N\sum_{m=1}^MR_k^n[m]}{\geq\gamma_k,}{\label{a_2}}
\end{maxi!}
where $\mathbf{x} = \{x_k^n[m]\}$ is the set of scheduling indicators.

To solve (\ref{eq:opt_2}), we introduce lack variables $d_k^n[m]$ and $h_k^n[m]$ to transform (\ref{eq:objectiveopt_2}) into
\[\sum_{m=1}^{M}\sum_{k=1}^{K}\sum_{n=1}^{N}\log_2(2^{d_k^n[m]}/2^{h_k^n[m]}) = \sum_{m=1}^{M}\sum_{k=1}^{K}\sum_{n=1}^{N} d_k^n[m] - h_k^n[m].\] Then, (\ref{eq:opt_2}) can be transformed into a mixed integer non-linear problem (MINLP):
\begin{maxi!}
{\mathbf{x}}{{\qquad \sum_{m=1}^{M}\sum_{k=1}^{K}\sum_{n=1}^{N} d_k^n[m] - h_k^n[m]} \label{eq:objectiveopt_minp}}{\label{eq:opt_minp}}{}
\addConstraint{(\ref{sche_cons_1})-}{(\ref{sche_cons_4})}{\label{s_minp}}
\addConstraint{\sum_{n=1}^N\sum_{m=1}^M d_k^n[m] - h_k^n[m]}{\geq\gamma_k}{\label{a_minp}}
\addConstraint{2^{d_k^n[m]}}{\leq SI_k^n[m]}{\label{b_minp}}
\addConstraint{2^{h_k^n[m]}}{\geq \sum_{n'\neq n, n'= 1}^N\sum_{k'\neq k, k'=1}^K Pc_{k',k}^{n'}[m]x_{k'}^{n'}[m] + \sigma^2,}{\label{c_minp}}
\end{maxi!}
where $SI_k^n[m]\triangleq Pg_{k}^n[m]x_k^n[m]+\sum_{n'\neq n, n'= 1}^N\sum_{k'\neq k, k'=1}^K Pc_{k',k}^{n'}x_{k'}^{n'}[m] + \sigma^2$.

Constraint (\ref{c_minp}) is not convex. To manage the non-convexity of the problem, we use the linear relaxation method, in which we find a linear upper bound of $2^{h_k^n[m]}$. To be specific, the lower bound for $h_k^n[m]$ is defined by $l_{h_k^n[m]} \triangleq \log_2(\sigma^2)$ and the upper bound for $h_k^n[m]$ is defined by
\begin{equation}
u_{h_k^n[m]} \triangleq \log_2\left(\sum_{n'\neq n, n'= 1}^N\sum_{k'\neq k, k'=1}^K Pc_{k',k}^{n'} + \sigma^2\right).    
\end{equation}
Then, the linear upper bound for $2^{h_k^n[m]}$ is 
\begin{equation}
\label{lp}
    2^{h_k^n[m]} \leq \frac{2^{u_{h_k^n[m]}} - 2^{l_{h_k^n[m]}}}{u_{h_k^n[m]}- l_{h_k^n[m]}}h_k^n[m] - \frac{l_{h_k^n[m]}2^{u_{h_k^n[m]}} - u_{h_k^n[m]}2^{l_{h_k^n[m]}}}{u_{h_k^n[m]}- l_{h_k^n[m]}}. 
\end{equation}
Therefore, (\ref{c_minp}) can be relaxed as
\begin{equation}
\begin{aligned}
\label{linar_relax}
    &\sum_{n'\neq n, n'= 1}^N\sum_{k'\neq k, k'=1}^K Pc_{k',k}^{n'}x_{k'}^{n'}[m] + \sigma^2\leq\\ &\frac{2^{u_{h_k^n[m]}} - 2^{l_{h_k^n[m]}}}{u_{h_k^n[m]}- l_{h_k^n[m]}}h_k^n[m] - \frac{l_{h_k^n[m]}2^{u_{h_k^n[m]}} - u_{h_k^n[m]}2^{l_{h_k^n[m]}}}{u_{h_k^n[m]}- l_{h_k^n[m]}},
\end{aligned}
\end{equation}
which converts the constraint optimization problem in (\ref{eq:opt_minp}) to
\begin{maxi!}
{\mathbf{x}}{{\qquad \sum_{m=1}^{M}\sum_{k=1}^{K}\sum_{n=1}^{N} d_k^n[m] - h_k^n[m]} \label{eq:objectiveopt_minp2}}{\label{eq:opt_minp2}}{}
\addConstraint{(\ref{sche_cons_1})-}{(\ref{sche_cons_4})}{\label{s_minp2}}
\addConstraint{\sum_{n=1}^N\sum_{m=1}^M d_k^n[m] - h_k^n[m]}{\geq\gamma_k}{\label{a_minp2}}
\addConstraint{2^{d_k^n[m]}}{\leq SI_d^k[n]}{\label{b_minp2}}
\addConstraint{(\ref{linar_relax}).}{}{\label{c_minp2}}
\end{maxi!}
To solve (\ref{eq:opt_minp2}), we will use the spatial branch-and-bound (sBnB) method \cite{belotti2013mixed}, which recursively partitions the feasible set. In general, sBnB searches a tree whose nodes correspond to sub-problems of the integer relaxation of (\ref{eq:opt_minp2}), i.e., (\ref{eq:opt_minp2}) without integer constraints, and whose edges correspond to branching decisions. We use both optimality and feasibility of subproblems to prune nodes in the tree.

A node in the branch-and-bound tree is uniquely defined by a set of
bounds, $(\mathbf{l}, \mathbf{u})$, on the integer variables and corresponds to (\ref{eq:opt_minp2}):
\begin{maxi!}
{\mathbf{x}}{{\qquad \sum_{m=1}^{M}\sum_{k=1}^{K}\sum_{n=1}^{N} d_k^n[m] - h_k^n[m]} \label{eq:objectiveopt_minp3}}{\label{eq:opt_minp3}}{}
\addConstraint{\sum_{n=1}^N\sum_{m=1}^M d_k^n[m] - h_k^n[m]}{\geq\gamma_k}{\label{a_minp2}}
\addConstraint{2^{d_k^n[m]}}{\leq SI_k^n[m]}{\label{b_minp3}}
\addConstraint{l_k^n[m] \leq x_k^n[m]}{\leq u_k^n[m]}{\label{bb}}
\addConstraint{(\ref{linar_relax}).}{}{\label{c_minp3}}
\end{maxi!}
We call (\ref{eq:opt_minp3}) the non-linear problem $\textbf{NLP}(\mathbf{l}, \mathbf{u})$. The root node of sBnB corresponds to $\textbf{NLP}(\mathbf{l}^0, \mathbf{u}^{0})$, where $\mathbf{l}^0 \triangleq [0,...,0]$ and $\mathbf{u}^0 \triangleq [1,...,1]$.
When we find a solution $\mathbf{x}'$ for $\textbf{NLP}(\mathbf{l}, \mathbf{u})$, there are two scenarios: (a) $\mathbf{x}'$ is feasible for (\ref{eq:opt_minp}). There are two possibilities in this scenario:
\begin{itemize}
    \item $\mathbf{x}'$ is the optimal solution for (\ref{eq:opt_minp}) and is accepted.
    \item $\mathbf{x}'$ is not optimal for (\ref{eq:opt_minp}) and is discarded.
\end{itemize}
In both cases, the subproblem $\textbf{NLP}(\mathbf{l}, \mathbf{u})$ can be eliminated, i.e., we will not perform branching on this node; (b) $\mathbf{x}'$ is
infeasible for (\ref{eq:opt_minp}), then at least one of the following two cases holds:
\begin{itemize}
    \item $\mathbf{x}'$ is not integer feasible, i.e., there exists $x_k^n[m] \neq 0$ or $x_k^n[m]' \neq 1$.
    \item Constraint (\ref{c_minp}) is violated.
\end{itemize}
In the first case, one can generate two new subproblems $\textbf{NLP}(\mathbf{l}^{-}, \mathbf{u}^{-})$ and
$\textbf{NLP}(\mathbf{l}^{+}, \mathbf{u}^{+})$, whose feasible regions $F(\mathbf{l}^{-}, \mathbf{u}^{-})$ and $F(\mathbf{l}^{+}, \mathbf{u}^{+})$ are created by the branching rule $x_k^n[m] \leq \lfloor x_k^n[m]'\rfloor \vee x_k^n[m] \geq \lceil{x_k^n[m]'}\rceil$, i.e., $x_k^n[m] = 0 \vee x_k^n[m] = 1$. In the second
case, branching may be necessary on a continuous variable. For example, if $h_k^n[m]'$ violates (\ref{c_minp}), then we can branch on $h_k^n[m]$ through $l_{h_k^n[m]} \leq  h_k^n[m] \leq h_k^n[m]' \vee h_k^n[m]' \leq  h_k^n[m] \leq u_{h_k^n[m]}$. On each branch, we also refine the constraint (\ref{linar_relax}) by the new boundaries of  $h_k^n[m]$. We call the nodes on each branch  $\hat{\textbf{NLP}}(\mathbf{l}^{+}, \mathbf{u}^{+})$ and $\hat{\textbf{NLP}}(\mathbf{l}^{-}, \mathbf{u}^{-})$. The details of the sBnB method is described in Alg. \ref{sBnB}. 
\begin{algorithm}[h]
\small
\caption{Spatial Branch-and-Bound}\label{sBnB}
\begin{algorithmic}[1]
\State Input the accuracy $\epsilon > 0$, set $U = -\infty$, and initialize the heap of the sBnB tree $\mathcal{H} = \varnothing$.
\State Add $\textbf{NLP}(\mathbf{l}^0, \mathbf{u}^{0})$ to the heap: $\mathcal{H} = \mathcal{H}\cup\{\textbf{NLP}(\mathbf{l}^0, \mathbf{u}^{0})\}$.
\While{$\mathcal{H}\neq\varnothing$}
\State Remove a node $\textbf{NLP}(\mathbf{l}, \mathbf{u})$ from the heap: $\mathcal{H} = \mathcal{H} - \{\textbf{NLP}(\mathbf{l}, \mathbf{u})\}$.
\State Solve $\textbf{NLP}(\mathbf{l}, \mathbf{u})$ and get the solution $\mathbf{x}^{(\mathbf{l}, \mathbf{u})}$.
\If{$\textbf{NLP}(\mathbf{l}, \mathbf{u})$ is infeasible}
\State Node can be pruned because it is infeasible.
\ElsIf{Optimal value of $\textbf{NLP}(\mathbf{l}, \mathbf{u})$ is less than $U$}
\State Node can be pruned, because it is dominated by lower bound.
\ElsIf{$\mathbf{x}^{(\mathbf{l}, \mathbf{u})}$ is integral and (\ref{c_minp}) is satisfied}
\State Update $U$ by the optimal value of $\textbf{NLP}(\mathbf{l}, \mathbf{u})$; $\mathbf{x}^{*} = \mathbf{x}^{(\mathbf{l}, \mathbf{u})}$.
\Else{}
\State Branch on Variable $(\mathbf{x}^{(\mathbf{l}, \mathbf{u})}, \mathbf{h}^{(\mathbf{l}, \mathbf{u})},\mathbf{l},\mathbf{u},\mathcal{H})$; See Alg. \ref{branch}.
\EndIf
\EndWhile
\end{algorithmic}
\end{algorithm}
\begin{algorithm}[h]
\small
\caption{Branch on Variable $(\mathbf{x}^{(\mathbf{l}, \mathbf{u})}, \mathbf{h}^{(\mathbf{l}, \mathbf{u})},\mathbf{l},\mathbf{u},\mathcal{H})$}\label{branch}
\begin{algorithmic}[1]
\If{$x_k^n[m]$ is integer infeasible}
\State Set $u_{x_k^n[m]}^{-} = \lfloor x_k^n[m]\rfloor$, $\mathbf{l}^{-} = \mathbf{l}$ and $l_{x_k^n[m]}^{+} = \lceil x_k^n[m]\rceil$, $\mathbf{u}^{+} = \mathbf{u}$.
\State Add $\hat{\textbf{NLP}}(\mathbf{l}^{-},\mathbf{u}^{-})$ and $\hat{\textbf{NLP}}(\mathbf{l}^{+},\mathbf{u}^{+})$ to the heap:
$\mathcal{H} = \mathcal{H} \cup \{\hat{\textbf{NLP}}(\mathbf{l}^{-},\mathbf{u}^{-}),\hat{\textbf{NLP}}(\mathbf{l}^{+},\mathbf{u}^{+})\}$.
\ElsIf{$h_k^n[m]$ violates (\ref{c_minp})}
\State Set $u_{h_k^n[m]}^{-} = h_k^n[m]$, $\mathbf{l}^{-} = \mathbf{l}$ and $l_{h_k^n[m]}^{+} = h_k^n[m]$, $\mathbf{u}^{+} = \mathbf{u}$.
\State Add $\hat{\textbf{NLP}}(\mathbf{l}^{-},\mathbf{u}^{-})$ and $\hat{\textbf{NLP}}(\mathbf{l}^{+},\mathbf{u}^{+})$ to the heap:
$\mathcal{H} = \mathcal{H} \cup \{\hat{\textbf{NLP}}(\mathbf{l}^{-},\mathbf{u}^{-}),\hat{\textbf{NLP}}(\mathbf{l}^{+},\mathbf{u}^{+})\}$
\EndIf
\end{algorithmic}
\end{algorithm}

\subsection{Deployment}
When optimizing the deployment, we fix the scheduling, the beamforming vector, and the RIS phases. We denote the index of the scheduled user by UAV $n$ at Timeslot $m$ by $i_m^n$, then the sub-problem for deployment can be expressed as
\begin{maxi!}
{\{\mathbf{p}\}}{{\qquad \sum_{m=1}^{M}\sum_{n=1}^N R_{i_m^n}} \label{eq:objectiveopt_deploy}}{\label{eq:opt_deploy}}{}
\addConstraint{|\mathbf{p}_n - \mathbf{p}^\text{pre}_n|}{= d\;\text{or}\; 0}{\label{b_deployment}}
\addConstraint{R_{i_m}^n}{\geq \gamma_{i_m^n}} {\label{a_deployment},}
\end{maxi!}
where $\gamma_{i_m^n}$ is calculated by $\gamma_{i_m}/(\sum_{m=1}^M\sum_{n=1}^N x_{i_m^n}^n[m])$ and $\{\mathbf{p}\}\triangleq\{\mathbf{p}_1,...,\mathbf{p}_N\}$.

The position of the moved UAV can be expressed as
\begin{equation}
\label{p_update}
    \mathbf{p}_n = \mathbf{p}_n^\text{pre} + d [\sin{\theta_n^\text{mv}}\cos{\varphi_n^\text{mv}}, \sin{\theta_n^\text{mv}}\sin{\varphi_n^\text{mv}}, \cos{\theta_n^\text{mv}}]^T,
\end{equation}
where $\theta_n^\text{mv}$ and $\varphi_n^\text{mv}$ are the movement elevation angle and azimuth from $\mathbf{p}_n$ to $\mathbf{p}_n^\text{pre}$, respectively. To find the best position for the UAV, we perform a sphere search for $\mathbf{p}_n$ based on $\mathbf{p}_n^\text{pre}$, i.e., we find the optimal moving direction from $\mathbf{p}_n^\text{pre}$ to $\mathbf{p}_n$. The detailed algorithm is described in Alg. \ref{dep_alg}.

\begin{algorithm}[h]
\small
\caption{Best deployment}\label{dep_alg}
\begin{algorithmic}[1]
\State \textbf{Input}: 
\State Searching step size $\Delta$, $\{\mathbf{p}_\text{pre}\}$ and the sum-rate of the previous timeblock $R_\text{sum}^\text{pre}$;
\State $\{\mathbf{p}^\text{opt}\} \gets \{\mathbf{p}_\text{pre}\}$, $R_\text{sum}^\text{max} \gets R_\text{sum}^\text{pre}$;
\For{$n = 1:N$}
\For{$\theta_\text{mv} = 0 : \Delta: 2\pi$}
\For{$\varphi_\text{mv} = 0:\Delta:2\pi$}
\State Update $\mathbf{p}_n$ and $R_\text{sum}$;
\If{$R_\text{sum} \geq R_\text{sum}^\text{max}$ and Constraint (\ref{a_deployment}) is satisfied}
\State \textbf{$\mathbf{p}_n^\text{opt} \gets \mathbf{p}_n$, $R_\text{sum}^\text{max} \gets R_\text{sum}$};
\EndIf
\EndFor
\EndFor
\EndFor
\State Return the optimal $\{\mathbf{p}^\text{opt}\}$.
\end{algorithmic}
\end{algorithm}

\subsection{Beamforming vector and RIS-phase design}
Given the scheduling order and the optimal deployment, we can simplify the problem of beamforming vector and RIS-phase design into
\begin{maxi!}
{\{\mathbf{w}_k^n\}, \{\mathbf{\Theta_r}\}}{{\qquad \sum_{m=1}^{M}\sum_{n=1}^{N} R_{i_m^n}} \label{eq:objectiveopt_3}}{\label{eq:opt_3}}{}
\addConstraint{R_{i_m^n}}{\geq \gamma}{\label{a_3}}
\addConstraint{|[\mathbf{w}_k^n[m]]_t|}{ = \frac{1}{\sqrt{N_t}}, t = 1,...,N_t, n = 1,...,N, k = 1,...,K}{\label{b_3}}
\addConstraint{\Theta_r[m]}{ = \text{diag}(e^{j\theta_{r,1,m}},...,e^{j\theta_{r,N_\text{RIS},m}}), r = 1,...,R,}{\label{c_3}}
\end{maxi!}
where we denote the index of scheduled user at Timeslot m for UAV $n$ as $i_m^n$. Since we already guarantee the minimum rate in the deployment and scheduling, we will discard the minimum rate constraint in the beamforming and RIS design.

To further simplify the problem, we decouple Problem (\ref{eq:opt_3}) by slots.
At Timeslot $m$, we need to design the beamforming vector for scheduled users at Timeslot $m$ and the RIS phases according to 
\begin{maxi!}
{\{\mathbf{w}_k^n\}, \{\mathbf{\Theta_r}\}}{{\qquad \sum_{n=1}^{N} R_{i_m^n}} \label{eq:objectiveopt_4}}{\label{eq:opt_4}}{}
%\addConstraint{R_{i_m^n}}{\geq \gamma}{\label{a_4}}
\addConstraint{|[\mathbf{w}_{i_m^n}^n]_t|}{ = \frac{1}{\sqrt{N_t}}, t = 1,...,N_t ,n = 1,...,N, i_m^n \in C_m}{\label{b_4}}
\addConstraint{\Theta_r[m]}{ = \text{diag}(e^{j\theta_{r,1,m}},...,e^{j\theta_{r,N_\text{RIS},m}}), r = 1,...,R.}{\label{c_4}}
\end{maxi!}
Problem (\ref{eq:opt_4}) is still intractable. Fortunately, according to \cite{christensen2008weighted}, there is a relationship between the data rate and the minimum mean-square error (MMSE). Since we only have one receiving antenna, the relationship can be expressed as
\begin{equation}
\label{relation}
    R_{i_m^n} = \log_2\left(\frac{1}{E_{i_m^n}^{\text{MMSE}}}\right),
\end{equation}
where 
\[
E_{i_m^n}^\text{MMSE} = \min(\mathrm{E}[(\tilde{s}_{i_m^n} - s_{i_m^n})(\tilde{s}_{i_m^n} - s_{i_m^n})^H]
\]
and 
\[\tilde{s}_{i_m^n} = u_{i_m^n}y_{i_m^n}.
\]
\[
y_{i_m^n} = \sqrt{P}(\mathbf{h}_{i_m^n}^{n^H} + \sum_{r=1}^R\mathbf{h}_{i_m^n}^r\mathbf{\Theta}_r[m]\mathbf{G}_r^n)\mathbf{w}_{i_m^n}^ns_{i_m^n} + \sum_{n'\neq n} \sqrt{P}(\mathbf{h}_{i_m^n}^{n'^H} + \sum_{r=1}^R\mathbf{h}_{i_m^n}^r\mathbf{\Theta}_r[m]\mathbf{G}_r^{n'})\mathbf{w}_{i_m^{n'}}^{n'}s_{i_m^{n'}} + n_{i_m^n}.
\]
$u_{i_m^n}\in C^{1\times 1}$ is a receiving weight for the estimated signal vector of each user.
Then, based on (\ref{relation}), maximizing the objective in (\ref{eq:opt_4}) is equivalent to minimizing the objective
\begin{equation}
\label{min_wmmse}
    \sum_{n=1}^N l_{i_m^n},
\end{equation}
where $l_{i_m^n} \triangleq \text{Tr}(g_{i_m^n}E_{i_m^n})-\log(g_{i_m^n})$, $E_{i_m^n} = \mathrm{E}[(\tilde{s}_{i_m^n} - s_{i_m^n})(\tilde{s}_{i_m^n} - s_{i_m^n})^H]$ is the mean square error (MSE). We call the minimization of Eq. (\ref{min_wmmse}) sum-WMMSE. Auxiliary variable $g_{i_m^n}$ functions as a weight. The equivalence between the objective in (\ref{eq:opt_4}) and (\ref{min_wmmse}) can be proved using the same method in \cite{christensen2008weighted}. We skip the detailed proof here for brevity. 

Based on the equivalence, Problem (\ref{eq:opt_4}) can be reformulated to 
\begin{mini!}
{\{g_{i_m^n}\},\{u_{i_m^n}\},\{\mathbf{w}_{i_m^n}^n\}, \{\mathbf{\Theta_r}\}}{{\qquad \sum_{n=1}^{N} l_{i_m^n}} \label{eq:objectiveopt_5}}{\label{eq:opt_5}}{}
%\addConstraint{R_{i_m^n}}{\geq \gamma_{i_m^n}}{\label{a_5}}
\addConstraint{|[\mathbf{w}_{i_m^n}^n]_t|}{ = \frac{1}{\sqrt{N_t}}, t = 1,...,N_t ,n = 1,...,N, i_m^n \in C_m}{\label{b_5}}
\addConstraint{\Theta_r[m]}{ = \text{diag}(e^{j\theta_{r,1,m}},...,e^{j\theta_{r,N_\text{RIS},m}}), r = 1,...,R}.{\label{c_5}}
\end{mini!}

% According to (\ref{relation}), noting that $s_{i_m^n} \in C^{1\times 1}$, $E_{i_m^n}^\text{MMSE} = \text{Tr}(E_{i_m^n}^\text{MMSE})$,
% Constraint (\ref{a_5}) can be reformulated as
% \begin{equation}
% \label{a_5_reform}
%     E_{i_m^n}^\text{MMSE} = \text{Tr}(E_{i_m^n}^\text{MMSE})\leq 2^{-\gamma_{i_m^n}}.
% \end{equation}
% Since $\text{Tr}(E_{i_m^n}^\text{MMSE}) \leq \text{Tr}(E_{i_m^n})$, we can further transform (\ref{a_5_reform}) to
% \begin{equation}
% \label{a_6_relax}
%     \text{Tr}(E_{i_m^n})\leq 2^{-\gamma_{i_m^n}}.
% \end{equation}

To solve Problem (\ref{eq:opt_5}), we will iterate among $\{\mathbf{W}_{i_m^n}\}$, $\{\mathbf{u}_{i_m^n}\}$, $\{\mathbf{w}_{i_m^n}^n\}$, and $\{\mathbf{\Theta_r}\}$ until convergence.

Note that $g_{i_m^n}$ does not appear in the constraints and we can directly derive the optimal solution for $g_{i_m^n}$ when $u_{i_m^n}$, $\mathbf{w}_{i_m^n}^n$, and $\mathbf{\Theta_r}$ are fixed. The optimal $g_{i_m^n}$ is given by
\begin{equation}
\label{W_imn}
    g_{i_m^n}^\text{opt} = \frac{1}{E_{i_m^n}},
\end{equation}
where $E_{i_m^n}$ is
\begin{equation}
\begin{aligned}
\label{E_imn}
E_{i_m^n} = (\sqrt{P}u_{i_m^n}\tilde{\mathbf{h}}_{i_m^n}^{n^H}\mathbf{w}_{i_m^n}^n - 1)(\sqrt{P}u_{i_m^n}\tilde{\mathbf{h}}_{i_m^n}^{n^H}\mathbf{w}_{i_m^n}^n - 1)^* + \sum_{n'\neq n}Pu_{i_m^n}\tilde{\mathbf{h}}_{i_m^n}^{n^{'H}}\mathbf{w}_{i_m^{n'}}^{n'}\mathbf{w}_{i_m^{n'}}^{n^{'H}}\tilde{\mathbf{h}}_{i_m^n}^{n'}u_{i_m^n}^* + \sigma^2|u_{i_m^n}|^2. 
\end{aligned}
\end{equation}
In (\ref{E_imn}), we denote $\mathbf{h}_{i_m^n}^{n^H} + \sum_{r=1}^R\mathbf{h}_{i_m^n}^r\mathbf{\Theta}_r[m]\mathbf{G}_r^n$ as $\mathbf{\tilde{h}}_{i_m^n}^{n^H}$.

When $g_{i_m^n}$ is fixed, the objective in (\ref{eq:opt_5}) is simplified to $\min\sum_{n=1}^N\text{Tr}(g_{i_m^n}E_{i_m^n})$. Note that when optimizing the $u_{i_m^n}$, the objective can be decoupled into $\min\text{Tr}(g_{i_m^n}E_{i_m^n}),\forall n$. Since $g_{i_m^n}$ is a variable (not a matrix),  $\min\text{Tr}(g_{i_m^n}E_{i_m^n}),\ \forall n$ can be simplified to $\min\text{Tr}(E_{i_m^n}),\ \forall n$. In this case, the best $u_{i_m^n}$ can be derived by setting $\frac{\partial\text{Tr}(E_{i_m^n})}{\partial u_{i_m^n}} = 0$. The optimal $u_{i_m^n}$ can be expressed as
\begin{equation}
\label{opt_u}
    u_{i_m^n}^\text{opt} = \mathbf{w}_{i_m^n}^{n^H}\mathbf{\tilde{h}}_{i_m^n}^n(P\mathbf{\tilde{h}}_{i_m^n}^{n^H}\mathbf{w}_{i_m^n}^{n}\mathbf{w}_{i_m^n}^{n^H}\mathbf{\tilde{h}}_{i_m^n}^n+\tilde{\Phi}_{i_m^n}^n)^{-1},
\end{equation}
where $\mathbf{\tilde{h}}_{i_m^n}^{n^H}=\mathbf{h}_{i_m^n}^{n^H} + \sum_{r=1}^R\mathbf{h}_{i_m^n}^r\mathbf{\Theta}_r\mathbf{G}_r^n$. $\tilde{R}_{i_m^n}^n$ is the interference-plus-noise variance, which is calculated as
\begin{equation}
    \tilde{\Phi}_{i_m^n}^n = \sum_{n'\neq n} P\mathbf{\tilde{h}}_{i_m^n}^{n^{'H}}\mathbf{w}_{i_m^{n'}}^{n'}\mathbf{w}_{i_m^{n'}}^{n^{'H}}\mathbf{\tilde{h}}_{i_m^{n}}^{n'} + \sigma^2. 
\end{equation}

\subsection{Optimizing the analog beamforming vector}
After optimizing $u_{i_m^n}$, we will turn to optimize the beamforming vector and RIS phases. 
% Note that we will discard Constraint (\ref{a_5}), which is the upper-bound for $\text{Tr}(E_{i_m^n})$, in the beamforming vector and RIS phases optimization since we have already minimized the $\text{Tr}(E_{i_m^n})$.
We reformulate the sum-WMMSE problem (\ref{min_wmmse}) as follows
\begin{equation}
\begin{aligned}
\label{decouple}
    \min\sum_{n=1}^N\text{Tr}(g_{i_m^n}E_{i_m^n}) & = \min\sum_{n=1}^N( \mathbf{w}_{i_m^n}^{n^H}(P\sum_{k=1}^N|g_{i_m^n}u_{i_m^k}|^2\mathbf{\tilde{h}}_{i_m^k}^n\mathbf{\tilde{h}}_{i_m^k}^{n^H})\mathbf{w}_{i_m^n}^{n}\\
    &-\sqrt{P}g_{i_m^n}u_{i_m^n}\mathbf{\tilde{h}}_{i_m^n}^{n^H}\mathbf{w}_{i_m^n}^n - \sqrt{P}g_{i_m^n}u_{i_m^n}^*\mathbf{w}_{i_m^n}^{n^H}\mathbf{\tilde{h}}_{i_m^n}^{n} + |u_{i_m^n}|^2\sigma^2 + 1).
\end{aligned}
\end{equation}
In (\ref{decouple}), $P\sum_{k=1}^N|u_{i_m^k}|^2\mathbf{\tilde{h}}_{i_m^k}^n\mathbf{\tilde{h}}_{i_m^k}^{n^H}$ is a constant matrix, so we can decouple the optimization problem among UAVs, i.e., we can separately optimize $\mathbf{w}_{i_m^n}$ for each $n$.
By neglecting the constant terms, the decoupled optimization problem is as follows
\begin{equation}
\begin{aligned}
\label{bf_opt}
    &\min_{\mathbf{w}_{i_m^n}^{n}} f(\mathbf{w}_{i_m^n}^{n})\\
    &\text{subject to } |[\mathbf{w}_{i_m^n}^n]_t|= \frac{1}{\sqrt{N_t}}, t = 1,...,N_t,
\end{aligned}
\end{equation}
where $f(\mathbf{w}_{i_m^n}^{n}) =  \mathbf{w}_{i_m^n}^{n^H}\mathbf{A}\mathbf{w}_{i_m^n}^{n}-2\sqrt{P}\text{Re}(g_{i_m^k}^*u_{i_m^n}^*\mathbf{w}_{i_m^n}^{n^H}\mathbf{\tilde{h}}_{i_m^n}^{n})$ and $\mathbf{A} \triangleq P\sum_{k=1}^N|g_{i_m^k}u_{i_m^k}|^2\mathbf{\tilde{h}}_{i_m^k}^n\mathbf{\tilde{h}}_{i_m^k}^{n^H}$. 

To solve (\ref{bf_opt}), we can use the majority-minimization (MM) algorithm, which is proposed in \cite{sun2016majorization,pan2020multicell}. It is a method which sequentially solves (\ref{bf_opt}) by constructing a series of more tractable approximate sub-problems. The sub-problem provides an upper-bound of $f(\mathbf{w}_{i_m^n}^{n})$. Denoting the objective of the sub-problem at Iteration $t$ as $h(\mathbf{w}_{i_m^n}^{n}|\mathbf{w}_{i_m^n}^{n,[t-1]})$, it will have the following properties:
\begin{enumerate}
    \item $h(\mathbf{w}_{i_m^n}^{n,[t-1]}|\mathbf{w}_{i_m^n}^{n,[t-1]}) = f(\mathbf{w}_{i_m^n}^{n,[t-1]})$
    \item $\nabla_{\mathbf{w}_{i_m^n}^{n}}h(\mathbf{w}_{i_m^n}^{n}|\mathbf{w}_{i_m^n}^{n,[t-1]})|_{\mathbf{w}_{i_m^n}^{n} = \mathbf{w}_{i_m^n}^{n,[t-1]}} =  \nabla_{\mathbf{w}_{i_m^n}^{n}}f(\mathbf{w}_{i_m^n}^{n})|_{\mathbf{w}_{i_m^n}^{n} = \mathbf{w}_{i_m^n}^{n,[t-1]}} $
    \item $h(\mathbf{w}_{i_m^n}^{n}|\mathbf{w}_{i_m^n}^{n,[t-1]})\geq f(\mathbf{w}_{i_m^n}^{n}) $
\end{enumerate}

Using the same technique as \cite{pan2020multicell}, we can construct a sub-problem at Iteration $t$ as follows:
\begin{equation}
\begin{aligned}
\label{sub_opt}
    &\max_{\mathbf{w}_{i_m^n}^{n}} 2\text{Re}(\mathbf{w}_{i_m^n}^{n^H}\mathbf{q}^{[t-1]})\\
    &\text{subject to } |[\mathbf{w}_{i_m^n}^n]_t|= \frac{1}{\sqrt{N_t}}, t = 1,...,N_t,
\end{aligned}
\end{equation}
where $\mathbf{q}^{[t-1]} = (\lambda_\text{max}\mathbf{I}_{N_t} - \mathbf{A})\mathbf{w}_{i_m^n}^{n,[t-1]} + \sqrt{P}u_{i_m^n}^*\mathbf{h}_{i_m^n}^{n}$. $\lambda_\text{max}$ is the maximum eigenvalue of $\mathbf{A}$ and $\mathbf{w}_{i_m^n}^{n,[t-1]}$ is the the solution of the sub-problem at Iteration $t-1$. The optimal solution to (\ref{sub_opt}) is given by
\begin{equation}
   \mathbf{w}_{i_m^n}^{n,[t]} = \frac{1}{\sqrt{N_t}}e^{j\arg (\mathbf{q}^{[t-1]})}. 
\end{equation}
According to \cite{pan2020multicell}, the MM algorithm converges to the Karush-Kuhn-Tucker (KKT) point of Problem (\ref{bf_opt}). The detailed algorithm is presented in Alg. \ref{bfalg}.

\begin{algorithm}[h]
\small
\caption{Majority Minimization}\label{bfalg}
\begin{algorithmic}[1]
\State Input the accuracy $\epsilon$ and initialize a feasible $\mathbf{w}_{i_m^n}^{n,[0]}$ and $f(\mathbf{w}_{i_m^n}^{n,[0]})$
\State $t\gets 0$
\Repeat
\State $t\gets t+1$
\State Calculate $\mathbf{q}^{[t-1]} = (\lambda_\text{max}\mathbf{I}_{N_t} - \mathbf{A})\mathbf{w}_{i_m^n}^{n,[t-1]} + Pu_{i_m^n}^*\mathbf{h}_{i_m^n}^{n}$;
\State Update $\mathbf{w}_{i_m^n}^{n,[t]} = \frac{1}{\sqrt{N_t}}e^{j\arg (\mathbf{q}^{[t-1]})}$ 
\State Calculate $f(\mathbf{w}_{i_m^n}^{n,[t]})$
\Until{$\frac{|f(\mathbf{w}_{i_m^n}^{n,[t]}) -f(\mathbf{w}_{i_m^n}^{n,[t-1]})|}{f(\mathbf{w}_{i_m^n}^{n,[t]})} \leq \epsilon$}
\State Return $\mathbf{w}_{i_m^n}^{n,[t]}$.
\end{algorithmic}
\end{algorithm}
\subsection{Optimizing RIS phases}
\label{RIS_opt}
To optimize the RIS phases, we first need to simplify the expression. Defining $\mathbf{H}_{i_m^n}^\text{RIS} = [\mathbf{h}_{i_m^n}^{1},\cdots,\mathbf{h}_{i_m^n}^{R}]\in \mathcal{C}^{1\times RN_{RIS}}$,
$\mathbf{\Theta}[m] = \text{diag}(\Theta_1[m],\cdots,\Theta_R[m])\in\mathcal{C}^{RN_{RIS}\times RN_{RIS}}$, and $\mathbf{G}_\text{RIS}^{n} = [\mathbf{G}_1^{n^H},\cdots,\mathbf{G}_R^{n^H}]^H\in\mathcal{C}^{RN_{RIS}\times N_t}$, the channel between User $i_m^n$ and UAV $n$ can be simplified to
\begin{equation}
    \mathbf{\tilde{h}}_{i_m^n}^{n^H} = \mathbf{h}_{i_m^n}^{n^H} + \mathbf{H}_{i_m^n}^\text{RIS}\mathbf{\Theta}[m]\mathbf{G}_\text{RIS}^n.
\end{equation}

By denoting $\mathbf{w}_{i_m^n}^n\mathbf{w}_{i_m^n}^{n^H}$ by $\mathbf{W}_{i_m^n}^n$, we can re-organize $\text{Tr}(E_{i_m^n})$ as
\begin{equation}
\begin{aligned}
    &\text{Tr}(g_{i_m^n}E_{i_m^n})\\ &=\text{Tr}(|g_{i_m^n}u_{i_m^n}|^2\sum_{k=1}^N(\mathbf{H}_{i_m^n}^\text{RIS}\mathbf{\Theta}[m]\mathbf{G}_\text{RIS}^k\mathbf{W}_{i_m^k}^k\mathbf{G}_\text{RIS}^{k^H}\mathbf{\Theta}^H[m]\mathbf{H}_{i_m^n}^{\text{RIS}^H} + \mathbf{h}_{i_m^n}^{k^H}\mathbf{W}_{i_m^k}^k\mathbf{G}_\text{RIS}^{k^H}\mathbf{\Theta}^H[m]\mathbf{H}_{i_m^n}^{\text{RIS}^H}\\
    &+\mathbf{H}_{i_m^n}^\text{RIS}\mathbf{\Theta}[m]\mathbf{G}_\text{RIS}^k\mathbf{W}_{i_m^k}^k\mathbf{h}_{i_m^n}^{k})- g_{i_m^n}u_{i_m^n}\mathbf{H}_{i_m^n}^\text{RIS}\mathbf{\Theta}[m]\mathbf{G}_\text{RIS}^n\mathbf{w}_{i_m^n}^n-  g_{i_m^n}^*u_{i_m^n}^*\mathbf{w}_{i_m^n}^{n^H}\mathbf{G}_\text{RIS}^{n^H}\mathbf{\Theta}^H[m]\mathbf{H}_{i_m^n}^{\text{RIS}^H} + \text{Const}_{i_m^n})\\
    &= \text{Tr}(\mathbf{H}_{i_m^n}^\text{RIS}\mathbf{\Theta}[m]\mathbf{A}\mathbf{\Theta}^H[m]\mathbf{H}_{i_m^n}^{\text{RIS}^H}) + \text{Tr}( \mathbf{B}_{i_m^n}^H\mathbf{\Theta}^H[m]\mathbf{H}_{i_m^n}^{\text{RIS}^H}) + \text{Tr}( \mathbf{H}_{i_m^n}^\text{RIS}\mathbf{\Theta}[m]\mathbf{B}_{i_m^n}) + \text{Tr}(\text{Const}_{i_m^n})\\
    & = \text{Tr}(\mathbf{\Theta}[m]\mathbf{A}\mathbf{\Theta}^H[m]\mathbf{H}_{i_m^n}^{\text{RIS}^H}\mathbf{H}_{i_m^n}^\text{RIS}) +  \text{Tr}( \mathbf{H}_{i_m^n}^{\text{RIS}^H}\mathbf{B}_{i_m^n}^H\mathbf{\Theta}^H[m]) + \text{Tr}( \mathbf{\Theta}[m]\mathbf{B}_{i_m^n}\mathbf{H}_{i_m^n}^\text{RIS}) + \text{Tr}(\text{Const}_{i_m^n})\\
    & =  \text{Tr}(\mathbf{\Theta}[m]\mathbf{A}\mathbf{\Theta}^H[m]\mathbf{C}_{i_m^n}) +  \text{Tr}( \mathbf{D}_{i_m^n}^H\mathbf{\Theta}^H[m]) + \text{Tr}( \mathbf{\Theta}[m]\mathbf{D}_{i_m^n}) + \text{Tr}(\text{Const}_{i_m^n}),
\end{aligned}
\end{equation}
where $\text{Const}_{i_m^n}$ is a constant term which is independent of $\mathbf{\Theta}$, Matrix $\mathbf{A} = |g_{i_m^n}u_{i_m^n}|^2\sum_{k=1}^N\mathbf{G}_\text{RIS}^k\mathbf{W}_{i_m^k}^k\mathbf{G}_\text{RIS}^{k^H}$, $\mathbf{B}_{i_m^n} = |g_{i_m^n}u_{i_m^n}|^2\sum_{k=1}^N(\mathbf{G}_\text{RIS}^k\mathbf{W}_{i_m^k}^k\mathbf{h}_{i_m^n}^{k}) - g_{i_m^n}u_{i_m^n}\mathbf{G}_\text{RIS}^n\mathbf{w}_{i_m^n}^n$,  $\mathbf{C}_{i_m^n} = \mathbf{H}_{i_m^n}^{\text{RIS}^H}\mathbf{H}_{i_m^n}^\text{RIS}$, and $\mathbf{D}_{i_m^n} = \mathbf{B}_{i_m^n}\mathbf{H}_{i_m^n}^\text{RIS}$.

Then, by neglecting the constant term, the sum-MMSE objective can be reformulated as
\begin{equation}
 \begin{aligned}
\label{decouple_phase}
    \min\sum_{n=1}^N\text{Tr}(E_{i_m^n}) & = \min\sum_{n=1}^N\text{Tr}( \mathbf{\Theta}[m]\mathbf{A}\mathbf{\Theta}^H[m]\mathbf{C}_{i_m^n}) +  \text{Tr}( \mathbf{D}_{i_m^n}^H\mathbf{\Theta}^H[m]) + \text{Tr}( \mathbf{\Theta}[m]\mathbf{D}_{i_m^n})\\
    & = \min\text{Tr} (\mathbf{\Theta}[m]\mathbf{A}\mathbf{\Theta}^H[m]\mathbf{C}) +  \text{Tr}( \mathbf{D}^H\mathbf{\Theta}^H[m]) + \text{Tr}( \mathbf{\Theta}[m]\mathbf{D})\\
    & \overset{(a)}{=} \min \mathbf{v}_{\Theta_m}^H\mathbf{E}\mathbf{v}_{\Theta_m} + \mathbf{v}_{\Theta_m}^T\mathbf{d} + \mathbf{d}^H\mathbf{v}_{\Theta_m}^*,
\end{aligned} 
\end{equation}
where $\mathbf{C} = \sum_{n=1}^N \mathbf{C}_{i_m^n}$, $\mathbf{D} = \sum_{n=1}^N\mathbf{D}_{i_m^n}$, and $\mathbf{E} = \mathbf{A}\odot \mathbf{C}^T$. Vector $\mathbf{v}_{\Theta_m}$ is the collection of diagonal elements of $\mathbf{\Theta}[m]$ and vector $\mathbf{d}$ is the collection of diagonal elements of $\mathbf{D}$. Equation (a) is because of $\text{Tr} (\mathbf{\Theta}[m]\mathbf{A}\mathbf{\Theta}^H[m]\mathbf{C}) = \mathbf{v}_{\Theta_m}^H\mathbf{A}\odot \mathbf{C}^T\mathbf{v}_{\Theta_m}$ \cite{zhang2017matrix}, where $\odot$ represents the Hadamard product.

Then, the optimization of $\mathbf{\Theta}$ can be formulated as
\begin{equation}
\begin{aligned}
\label{phase_opt}
    &\min_{\mathbf{v}_{\Theta_m}} f(\mathbf{v}_{\Theta_m})\\
    &\text{subject to } |[\mathbf{v}_{\Theta_m}]_t|= 1, t = 1,...,RN_{RIS},
\end{aligned}
\end{equation}
where $f(\mathbf{v}_{\Theta_m}) = \mathbf{v}_{\Theta_m}^H\mathbf{E}\mathbf{v}_{\Theta_m}+2\text{Re}(\mathbf{v}_{\Theta_m}^H\mathbf{d}^*)$.

Using the MM algorithm, described for the analog beamforming design, we can find the solution to (\ref{phase_opt}).  The sub-problem at Iteration $t$ is
\begin{equation}
\begin{aligned}
\label{sub_opt_phase}
    &\max_{\mathbf{v}_{\Theta_m}} 2\text{Re}(\mathbf{v}_{\Theta_m}^{H}\mathbf{q}^{[t-1]})\\
    &\text{subject to } |[\mathbf{v}_{\Theta_m}]_t|= 1, t = 1,...,RN_{RIS},
\end{aligned}
\end{equation}
where $\mathbf{q}^{[t-1]} = (\lambda_\text{max}\mathbf{I}_{RN_{RIS}} - \mathbf{E})\mathbf{v}_{\Theta_m}^{[t-1]} + \mathbf{d}^*$. $\lambda_\text{max}$ is the maximum eigenvalue of $\mathbf{E}$ and $\mathbf{v}_{\Theta}^{[t-1]}$ is the solution of the sub-problem at Iteration $t-1$. The optimal solution to (\ref{sub_opt_phase}) is given by
\begin{equation}
   \mathbf{v}_{\Theta_m}^{[t]} = e^{j\arg \mathbf{q}^{[t-1]}}. 
\end{equation}
%Then, we can use Alg. (\ref{beam_opt}) to find the optimal phases.

We summarize the beamforming and RIS-phase optimization in Alg. \ref{beam_opt}. At each Timeslot $m$, we use Alg. \ref{beam_opt} to obtain the optimal beamforming vector and RIS phases.
\begin{algorithm}[h]
\caption{Joint beamforming and RIS-phase optimization}
\label{beam_opt}
\begin{algorithmic}[1]
\State Set sum-WMSE $M_{sum}[-1] \gets 0$, the maximal iteration number $k_{max} \gets 1000$, the convergence threshold $\epsilon \gets 10^{-3}$, and $k\gets 0$;
\State Randomly initialize the beamforming vector, the RIS phases, and the weight variable $g_{i_m^n}$;
\While{$|M_{sum}[k-1] - M_{sum}[k]| \geq \epsilon M_{sum}[k-1]$ and $k \leq k_{max}$}
\State Calculate the $u_{i_m^n}$ for User $i_m^n$ according to (\ref{opt_u});
\State Update $g_{i_m^n}$ for all $i_m^n$ according to (\ref{W_imn});
\State Fixing $u_{i_m^n}$ and RIS phases, optimize $\mathbf{w}_{i_m^n}$ for all $i_m^n$ using Alg.(\ref{bfalg});
\State Fixing $u_{i_m^n}$ and $\mathbf{w}_{i_m^n}$, optimize RIS phases $\mathbf{\Theta}[m]$ using the method in Sec. \ref{RIS_opt};
\EndWhile
\State Return $\{\mathbf{w}_{i_m^n}\}$ and $\mathbf{\Theta}[m]$.
\end{algorithmic}
\end{algorithm}
%\subsection{Complexity analysis of the joint optimization}

The details of the algorithm to jointly  optimize  deployment,  scheduling,  beamforming,  and  RIS  phases are described in Alg. \ref{joint}. Obviously, Alg. \ref{joint} converges since we generate a monotonically increasing sequence with an upper bound (the maximum sum-rate).
\begin{algorithm}[h]
\caption{Joint Optimization}\label{joint}
\begin{algorithmic}[1]
\State Set the sum-rate $R_\text{sum}[-1]\gets 0$, the maximal iteration number $k_{max}\gets 1000$, the convergence threshold $\epsilon\gets 10^{-3}$, and $k \gets 0$; 
\State Choose feasible start points $\mathbf{p}^\text{opt}[0]$, $\mathbf{x}^\text{opt}[0]$, $\{\mathbf{w}_k^{n,\text{opt}}[0]\}$, and $\mathbf{\Theta}^\text{opt}[0]$;
\While{$R_\text{sum}[k]-R_\text{sum}[k-1]\geq\epsilon R_\text{sum}[k-1]$ and $k\leq k_{max}$}
\State $k\gets k+1$;
\State Use Alg. \ref{dep_alg} to find the optimal deployment;
\State Obtain the optimal scheduling by using sBnB to solve (\ref{eq:opt_2});
\State Obtain the optimal beamforming vector and RIS phases by Alg. \ref{beam_opt};
\State Calculate $R_\text{sum}[k]$;
\EndWhile
\State Return $\mathbf{p}^\text{opt}$, $\mathbf{x}^\text{opt}$, $\{\mathbf{w}_k^{n,\text{opt}}\}$, and $\mathbf{\Theta}^\text{opt}$.
\end{algorithmic}
\end{algorithm}
\subsection{Complexity analysis}
In this section, we analyze the complexity of the optimization algorithm. 
For the deployment, the complexity is $\mathcal{O}(\frac{4\pi^2}{\Delta^2})$, where $\Delta$ is the step size of the sphere search.

For the sBnB method, the complexity is $\mathcal{O}(2^{KMN})$, where $K$ is the number of ground users, $M$ is the total number of timeslots, and $N$ is the number of UAVs.

For the joint beamforming and RIS-phase optimization algorithm, we iteratively optimize the beamforming vector and RIS phases. For both  beamforming optimization and RIS optimization, the MM algorithm is used. According to \cite{pan2020multicell}, the complexity of MM algorithm is $\mathcal{O}(M^3 + T_{MM}M^2)$, where $M$ is the number of optimization variables and $T_{MM}$ is the number of iterations required for the MM algorithm to converge. Then, for beamforming optimization, the complexity is $\mathcal{O}(N_t^3 + T_{MM}N_t^2)$, where $N_t$ is the number of antennas carried by one UAV. For the RIS-phase optimization, the complexity is $\mathcal{O}(R^3N_{RIS}^3 + T_{MM}R^2N_{RIS}^2)$, where $R$ is the number of RISs and $N_{RIS}$ is the number of reflecting elements in each RIS.

The overall complexity of the joint optimization algorithm is $\mathcal{O}(T_{Joint}(\frac{4\pi^2}{\Delta^2} + 2^{KMN} + N_t^3 + T_{MM}N_t^2 + R^3N_{RIS}^3 + T_{MM}R^2N_{RIS}^2))$, where $T_{Joint}$ is the number of iterations required for the joint optimization algorithm to converge.
The sBnB method is the complexity bottleneck of the joint optimization algorithm due to its exponential complexity. To reduce the complexity, one can fix the scheduling and optimize the other three components. Such an algorithm will have a polynomial complexity. We provide the comparison between the optimized scheduling and fixed scheduling in the simulation part. The optimized scheduling provides a moderate  sum-rate improvement (about $10\%$) compared to the fixed scheduling. Therefore, if some compromise can be made in terms of the sum-rate, the fixed scheduling algorithm is preferred because of its low polynomial complexity.  

\section{Simulation results}
\label{sec:simulation}
In this section, we provide simulation results for our proposed joint optimization algorithm. We consider a scenario where 2 UAVs serve 4 users in 4 timeslots with the assistance of 2 RISs. The UAVs serve the users using a mmWave carrier. We choose 28 GHz as the carrier's frequency, since 28 GHz is a typical frequency band in urban areas \cite{rappaport2015wideband}. The parameters in Eq. (\ref{plos}) are set as $a = 11.95$ and $b = 0.14$ \cite{al2014optimal}. The channel gain coefficient $a_k^n$ is generated according to a complex Gaussian distribution $a_k^n \sim \mathcal{CN}(0,10^{-0.1\kappa})$, where $\kappa = e + 10f\log_{10}(s) + \eta$. Parameter $s$ is the distance between the UAV and the user. We calculate $s$ according to the UAV's position in the previous timeblock. Parameters $f$ and $e$ are constants and $\eta \sim \mathcal{N}(0,\sigma_{\eta})$. For no-blockage scenario, $f = 2$, $e = 61.4$, and $\sigma_{\eta} = 5.8$. For blockage scenario, $f = 2.92$, $e = 72$, and $\sigma_{\eta} = 8.7$ \cite{rappaport2015wideband}. 

In our simulations, the RIS positions are $(25,25,0)$ and $(75,75,0)$. The UAVs and the RISs are all equipped with a 64 ($16\times 4$) antenna array. We set the amplitude of the moving step for the UAV to be $1$ meter. The initial positions of the UAVs are $(15,15,30)$ and $(35,35,30)$. We randomly generate the positions of the users and keep them throughout the simulation. The total number of timeblocks is 1000 and the channels change randomly from one timeblock to another, as explained in Section \ref{sec:channel}. We use the averaged sum-rate and minimum rate per timeblock to measure the performance of our system.
\begin{figure}[htb]
    \centering
    \includegraphics[width = 12 cm]{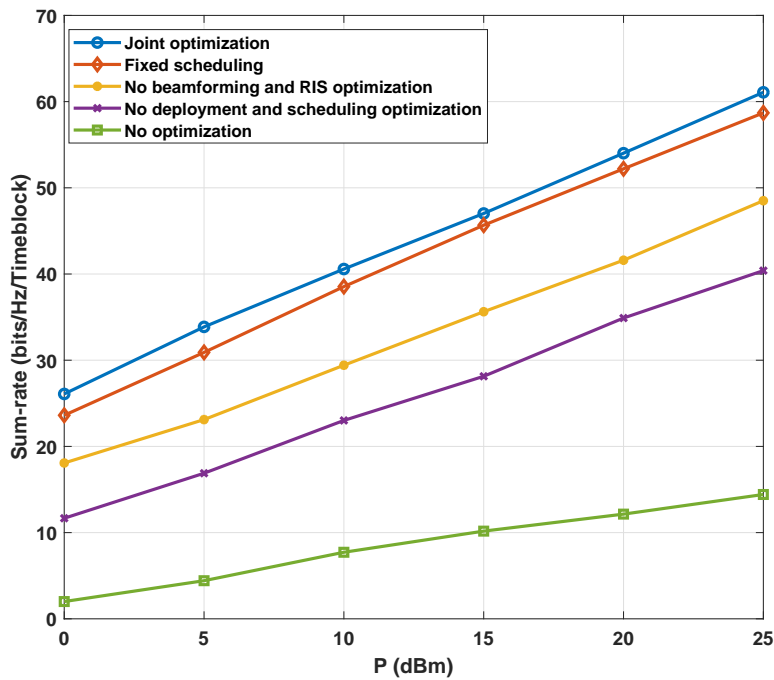}
    \caption{Sum-rate comparison}
    \label{sum_rates}
\end{figure}

\begin{figure}[htb]
    \centering
    \includegraphics[width = 12 cm]{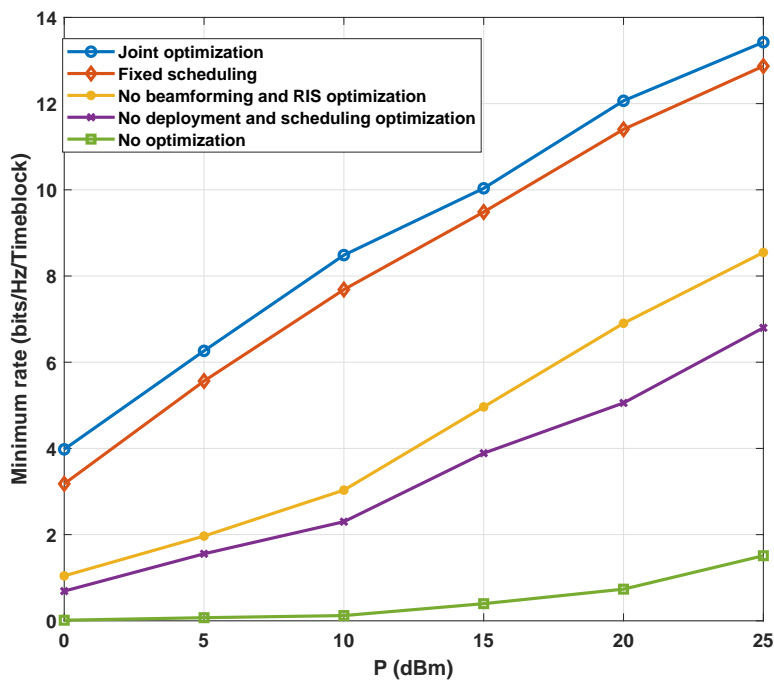}
    \caption{Minimum rate comparison}
    \label{min_rates}
\end{figure}
\begin{figure}[!htb]
    \centering
    \includegraphics[width = 12 cm]{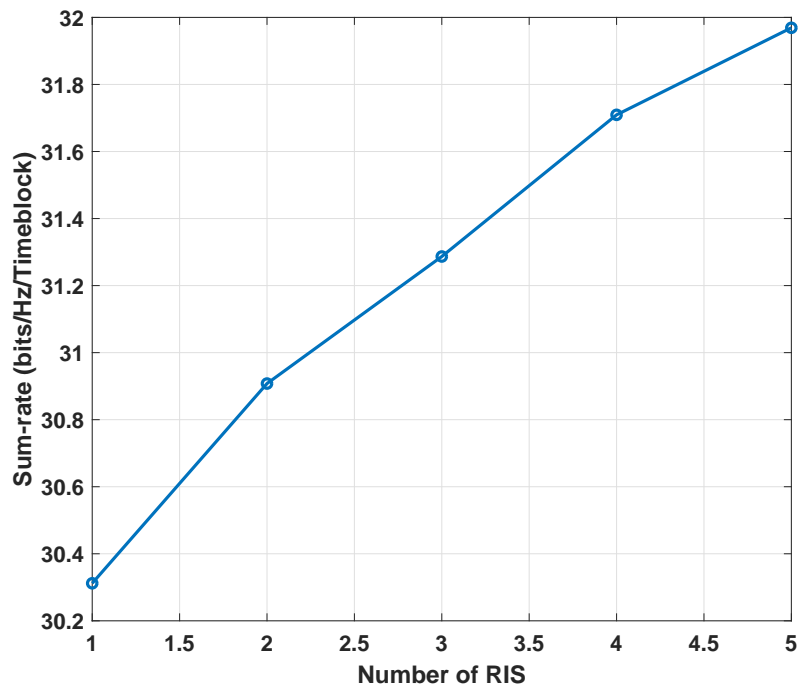}
    \caption{Sum-rate vs the number of RISs}
    \label{no_RIS}
\end{figure}

\begin{figure}[!htb]
    \centering
    \includegraphics[width = 12 cm]{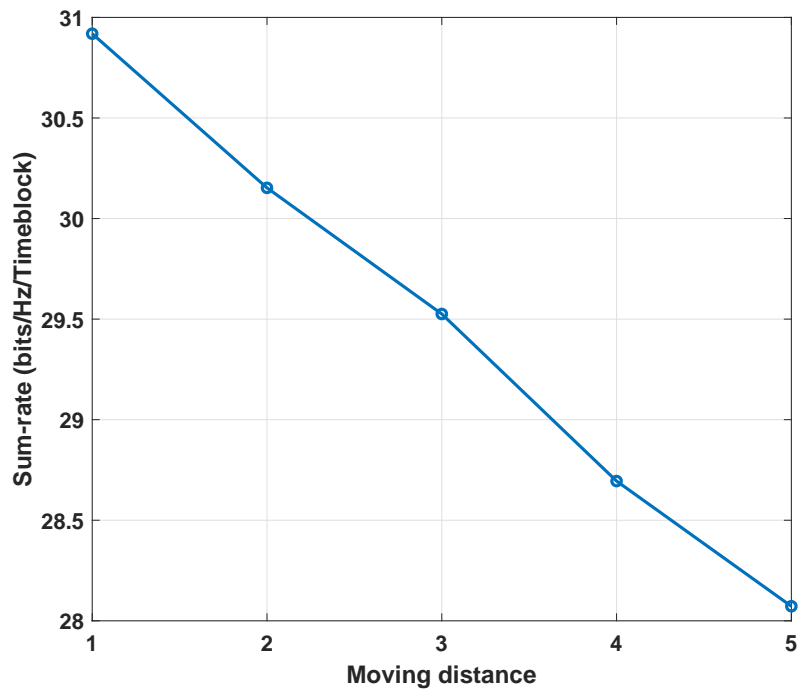}
    \caption{Sum-rate vs moving distance}
    \label{moving_distance}
\end{figure}

\begin{figure}[!htb]
    \centering
    \includegraphics[width = 12 cm]{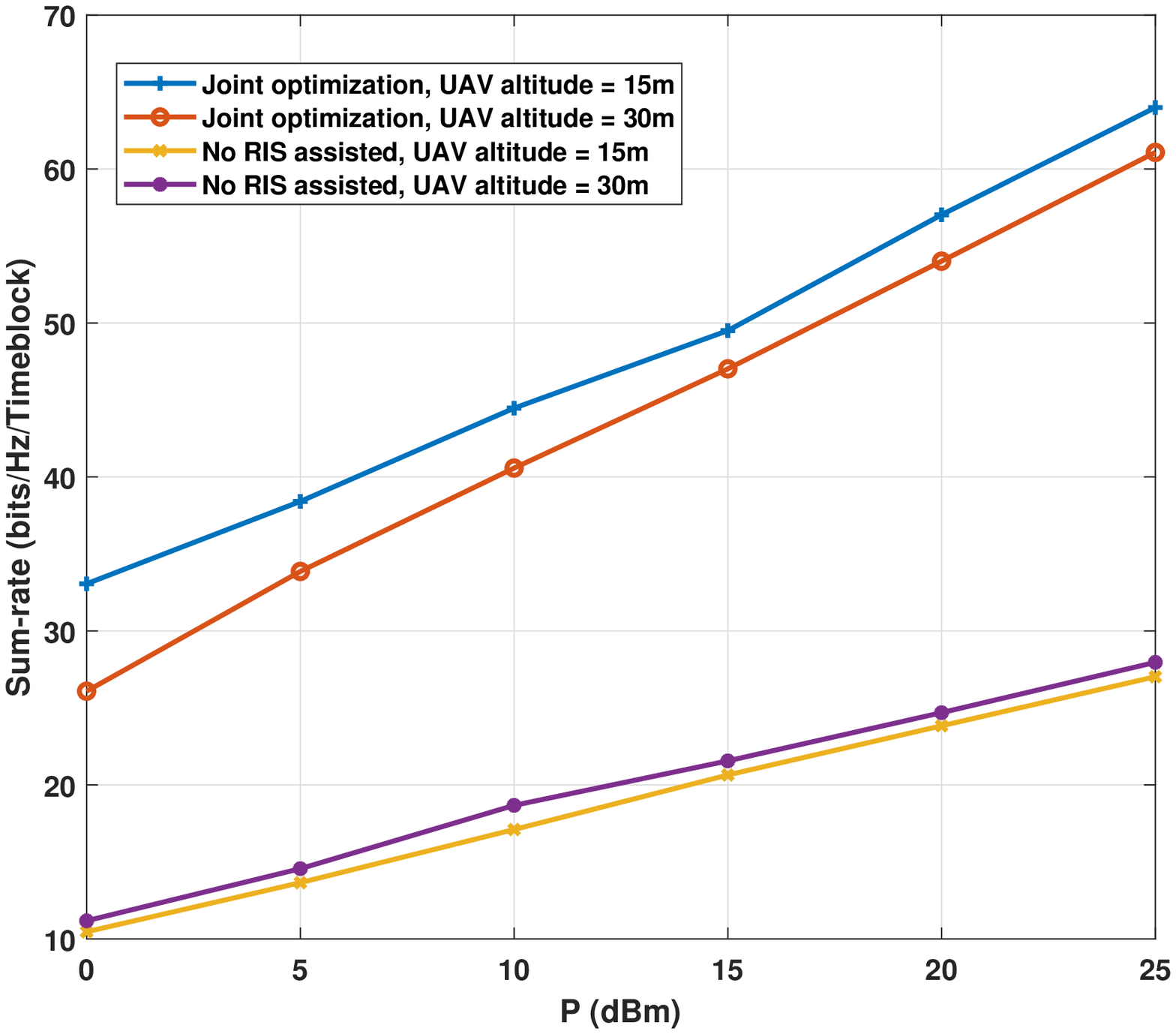}
    \caption{Effect of UAV altitude on sum-rate}
    \label{UAV_height}
\end{figure}

In Fig. \ref{sum_rates}, we compare the sum-rate among (i) the system which uses our proposed joint optimization method, (ii) the system using fixed scheduling, (iii) 
the system that does not optimize the beamforming and RIS, (iv) the system which does not optimize the deployment and scheduling, and (v) the system which uses the initial UAV location and randomly generates scheduling, beamforming vector, and RIS phases. The power of the Gaussian white noise is set to be $-85$ dBm and the minimum rate constraint is 1 bit/Hz/Timeblock. Note that for Systems (iv) and (v), there is no minimum rate constraint as it cannot be guaranteed without an appropriate deployment. The results show that our joint optimization method provides significant gains, in terms of the sum-rate, over the other systems. In particular, the joint optimization algorithm provides around $10\%$ gain over the fixed scheduling algorithm, which provides the closest performance to that of our system. It is worth mentioning that the dominant complexity component of the joint algorithm lies in the scheduling optimization. When complexity is a bottleneck, one can use the fixed scheduling to reduce the complexity with a relatively moderate loss in the sum-rate. Another observation is that the system with optimized deployment outperforms the system which only optimizes the beamforming vector and RIS phases. One can conclude that the beamforming and RIS optimization alone cannot compensate the performance loss brought by the bad channel conditions.

Fig. \ref{min_rates} compares the minimum rate among the same systems. %compared in Fig.  \ref{sum_rates}. %which uses our proposed joint optimization method, the system that only optimizes the deployment, the system which only optimizes the beamforming vector and RIS phases, and the system without any optimization. 
Our joint optimization method shows similar gains over the other four systems in terms of the minimum rate. Also, the performance order in Fig. \ref{min_rates} is the same as that of Fig.  \ref{sum_rates}.

In Fig. \ref{no_RIS}, we evaluate the system performance with different number of RISs. The locations of the RISs are chosen from the set $\{(5, 5, 0), (15, 15, 0), (25, 25, 0) (50, 50, 0), (75, 75, 0)\}$ randomly. The transmission power is set to be $5$ dBm. We use the fixed scheduling in the simulation. As the number of RISs increases, the system's sum-rate gradually increases. Note that, as the number of RISs increases, the implementation complexity increases too.

In Fig. \ref{moving_distance}, we evaluate the relationship between the system performance and the moving distance of the UAVs. The transmission power is set to be $5$ dBm. We use the fixed scheduling in the simulation. As the moving distance increases, the system performance gradually degrades. This is because a larger moving distance indicates a coarser resolution of the location search, which will degrade the performance of the deployment.

In Fig. \ref{UAV_height}, we compare the sum-rate of two systems at different altitudes: the system without RIS and the system using out joint optimization algorithm. For the system without RIS, we use random beamforming. As the UAV altitude decreases from $30$ meters to $15$ meters in the system without RIS, the sum-rate also decreases. The decrease is 
caused by the increase of the blockage probability. As pointed out in the channel model section, the blockage probability increases as the UAV altitude decreases. For the system using our joint optimization algorithm, as the UAV altitude decreases from $30$ meters to $15$ meters, there is an increase in the sum-rate because the path-loss is greatly reduced. Fig. \ref{UAV_height} shows the importance of using the RIS to compensate for the loss of throughput due to the blockage in low UAV altitude.

\section{Conclusions}
\label{sec:conclusion}
In this paper, we jointly optimized the deployment, user scheduling, beamforming vector, and RIS phases in a RIS-assisted UAV wireless network. To solve the problem, we iteratively optimized one of the four variables while fixed the other three variables.  For the deployment, we found the optimal position by a sphere search. Then, we formulated a MINLP to find the best scheduling and used an sBnB method to solve it. We also designed the analog beamforming vector and RIS phases using an iterative algorithm which makes use of the equivalent relationship between sum-rate maximization and sum-WMMSE. The proposed joint optimization outperforms the system using fixed scheduling, the system without beamforming and RIS optimization, and the system without deployment optimization.

\section*{Acknowledgement}
The authors would like to thank Mojtaba Ahmadi Almasi for helpful discussions.

\bibliographystyle{IEEEtran}
\bibliography{IEEEabrv,refs}
\end{document}